\documentstyle[epsf,12pt,psfig]{article}
\newcommand{\yourtitle}[1]{
\mbox{}\\
\vskip 4\baselineskip
{\bf\noindent #1}\\ }
\newcommand{\youraddress}[1]{
\noindent\mbox{}\hspace{1in}\parbox[t]{4.0in}{#1}\\ }
\newcommand{\yournames}[1]{
\mbox{}\\
\mbox{}\\
\noindent\mbox{}\hspace{1in}{#1}\\ }
\newcommand{\yourabstract}[1]{
\mbox{}\\
\mbox{}\\
{\bf\noindent Abstract}\\
\begin{center}
\mbox{}\parbox[t]{5.in}{#1}
\end{center} }
\newcommand{\yoursection}[1]{
\vskip 2\baselineskip
{\bf\noindent #1}\\
\mbox{}\\
\vspace{-0.19in}}
\newcommand{\referencestyle}
\small
\vspace{12pt}}

\begin{document}

\yourtitle{PARTON INTERACTION RATES AND LIMITS OF PERTURBATIVE QCD}
\yournames{Markus H. Thoma}\\
\youraddress{Institut f\"ur Theoretische Physik,\\
        Universit\"at Giessen, \\
        35392 Giessen, Germany}
\yourabstract{The relevance of parton interaction rates for observables
of a quark-gluon plasma and for testing field theoretical methods
is exhibited. Using the Braaten-Pisarski method the interaction rate
as well as the transport rate beyond the leading logarithm approximation
are estimated. Limits on the validity of
perturbative QCD for relativistic heavy ion collisions are discussed.}

\yoursection{INTRODUCTION}


The interaction (or damping) rates of parton scattering processes in a
thermalized QGP are interesting quantities. For example, considering
elastic scattering, $gg\rightarrow gg$, $gq\rightarrow gq$, $qq\rightarrow
qq$, the corresponding interaction rates $\Gamma $ are closely related to
the following important observables:

1. The relaxation time $\tau $, which provides information on the
thermalization of the parton gas in relativistic heavy ion collisions
\cite{SHA}, is just the inverse of the interaction rate, $\tau =
1/\Gamma $.

2. The mean free path $\lambda $ of a parton with velocity $v$
in the quark-gluon plasma (QGP) is given by $\lambda = v/\Gamma $.

3. The collisional energy loss $dE/dx$ [2-8] of a parton in the QGP
follows from $dE/dx = \Gamma \, \omega /v$, where $\omega $ is the
energy transfer per collision. The energy loss of a high energy parton
determines the amount of jet quenching in a relativistic heavy ion
collision, which might serve as a signature for the QGP \cite{GYP,GPT}.

4. The shear viscosity coefficient $\eta $ may be obtained from
$\eta = 2\epsilon /(15\Gamma )$ \cite{DAG}, where $\epsilon $ is the energy
density of the QGP. This quantity indicates the relevance of dissipation
in the expansion phase of the QGP.

5. The static color conductivity is related to the interaction
rate by $\sigma _c = 4m_g^2/\Gamma $ \cite{SGA}, where the effective
gluon mass or plasma frequency is given by $m_g^2=g^2T^2\, (1+N_f/6)/3$,
$N_f$ denoting the number of active flavors in the QGP.

Inelastic scattering rates, e.g. $gg\rightarrow ggg$, $gg\rightarrow
q\bar q$, on the other hand, are the inputs for the radiative energy loss
by gluon bremsstrahlung \cite{GPT,GWA} and describe the chemical
equilibration of the QGP \cite{BDM,XSA} .

Here, we will focus on the elastic rates only. We have to distinguish
between two kinds of rates, namely the interaction rate approximately
given by $\Gamma \simeq n\, \sigma $, where $n$ is the number density of
the partons in the QGP and $\sigma $ the cross section of the process
under consideration, and the {\it transport} interaction rate given
by $\Gamma _{trans} \simeq n\, \sigma _{trans}$ with the transport cross
section $\sigma _{trans} = \int d\sigma \, (\sin ^2\theta )/2$. Here, $\theta $
denotes the scattering angle in the center of mass system. The transport
weight $(\sin ^2\theta )/2$ suppresses collinear and anticollinear
scattering events, which are less important for momentum ralaxation
[16-19]. Thus $\Gamma _{trans}$ gives an estimate for
the thermalization time rather than $\Gamma $.

Since interaction rates are dynamical quantities which cannot be addressed
by lattice calculations, the only method for computing them is perturbation
theory. However, perturbative QCD (pQCD) at finite temperature suffers
from serious problems leading to infrared singularities and gauge
dependence for many observables calculated in this way. The reason for
this is the fact that naive perturbation theory at finite temperature
is incomplete, i.e. an expansion in the number of loops is not equivalent
to an expansion in the coupling constant $g$. In other words, multiple
loop diagrams may contribute to lower order in $g$ \cite{PIA}. This
problem can be circumvented by using effective propagators and vertices
based on a resummation of the so-called hard thermal loop diagrams, as
shown by Braaten and Pisarski \cite{BPA}. In this way, medium effects,
e.g. Debye screening, are included, improving the infrared behavior
of the results drastically. At the same time, the effective
perturbation theory leads to consistent results for observables, i.e.
gauge independent results which are complete to leading order in $g$.

However, although the effective perturbation theory means a crucial
improvement compared to the naive one, some problems still remain:

1. The effective propagators and vertices show a complicated
momentum and energy dependence, rendering explicit calculations
painful. Thus for many purposes a simplified version of the
Braaten-Pisarski method is desirable. A widely used
approximation consists in using naive perturbation theory but including
screening masses in the propagators. The interaction rates are ideal
examples for comparing this approximation with the effective perturbation
theory.

2. The effective gluon propagator does not contain a static magnetic
screening, leading to infrared singularities -- although less severe than
in naive perturbation theory -- in certain quantities, e.g. in the
ordinary interaction rate.

3. The effective perturbation theory has been derived under the
weak coupling limit assumption, $g\ll 1$, because it is based on a
separation between momentum scales $T$, $gT$, and $g^2T$. However,
realistic values of the strong coupling constant at relativistic heavy
ion collisions are expected to be of the order of $g \simeq 1.5$ -- 2.5
corresponding to $\alpha _s = g^2/(4\pi ) \simeq 0.2$ -- 0.5. In particular,
the transport rate and the energy loss provide limits for the extrapolation
to realistic values of $\alpha _s$, as will be discussed below.

Hence, the interaction rates not only determine important quantities of
the QGP but also can be used for studying essential questions of finite
temperature field theory and its applicability.

\yoursection{INTERACTION RATES}

Interaction rates can be calculated by starting either from matrix elements
or from self energies \cite{WEA}. The first way, for example in the case
of $qq\rightarrow qq$ scattering for massless quarks, gives
\begin{eqnarray}
\Gamma _{qq} & = & {1\over 2p}\> \int {d^3p' \over (2\pi )^3 2p'}\>
[1-n_F(p')]\> \int {d^3k\over (2\pi )^3 2k}\> n_F(k)\nonumber \\
&\times & \int {d^3k'\over (2\pi )^3 2k'}\> [1-n_F(k')]\> (2\pi )^4\>
\delta ^4(P+K-P'-K')\> 6N_f\> \langle |{\cal M}(qq\rightarrow qq)|^2
\rangle ,\nonumber \\
\label{eq1}
\end{eqnarray}
where $P = (p_0,{\bf p})$ denotes the four momentum and $n_f(p) =
1/[\exp (p/T)+1]$ with $p = |{\bf p}|$ the Fermi-Dirac distribution.
To lowest order pQCD the matrix element ${\cal M}$ is given by the
one-gluon exchange diagram shown in Fig.1a. The second way consists
in considering the imaginary part of the gluon self energy on mass
shell:
\begin{equation}
\Gamma _{qq}(p)=-{1\over 2p}\> [1-n_F(p)]\> tr\left [\gamma ^\mu P_\mu
\, Im \Sigma (p,{\bf p})\right ].
\label{eq2}
\end{equation}
where we have to take into account the self energy of Fig.1b. The equivalence
of the both methods follows from a generalization of the Cutkosky cutting
rules to finite temperatures \cite{KOS}.

\begin{figure}[htp]
\centerline{\psfig{figure=lbl1a.ps,height=10.0cm}
\hfill{\psfig{figure=lbl1b.ps,height=10.0cm}}}
\vskip-5cm
\caption{One-gluon exchange diagram and quark self energy entering the
quark interaction rate to lowest order naive perturbation theory.}
\end{figure}

At finite temperature the self energy of Fig.1b has to be computed by using
either the real or imaginary time formalism. The interaction rates
following from Fig.1a or Fig.1b turn out to be quadratically infrared
divergent
\begin{equation}
\Gamma \sim \alpha _s^2\> \int {dq\over q^3}.
\label{eq3}
\end{equation}
Using the Braaten-Pisarski method, derived for the imaginary time formalism,
we replace the self energy of Fig.1b by Fig.2. The effective gluon
propagator showing up there contains an infinite sum of gluon self
energy diagrams in the hard thermal loop approximation \cite{BPA},
which include an imaginary part. Assuming a hard momentum $p
{\buildrel > \over \sim } T$ for the external quark, the use of an
effective gluon propagator is sufficient for a consistent result \cite{PIA}.

\begin{figure}[htp]
\vskip-1.5cm
\centerline{\psfig{figure=lbl2.ps,height=20.0cm}}
\vskip-12.5cm
\caption{Quark self energy containing the effective gluon propagator
entering the quark interaction rate to lowest order effective
perturbation theory.}
\end{figure}

The final result reads [3, 17, 24-27]
\begin{eqnarray}
\Gamma _q^l & = & 1.098\> C_F\> \alpha _s\> T,\nonumber \\
\Gamma _q^t & \simeq  & C_F\> \alpha _s\> T\> \int {dq\over q},\nonumber \\
\label{eq4}
\end{eqnarray}
where $\Gamma _q^{l,t }$ denotes the part of the interaction rate
corresponding to the exchange of a longitudinal or transverse gluon
and $C_F=4/3$ is the Casimir invariant.

Due to the use of the resummed gluon propagator containing the effective
gluon mass proportional to $\alpha _s$ in the denominator, the interaction
rate (\ref{eq4}) is of lower order in $\alpha _s$ than expected naively
(\ref{eq3}). Furthermore, the rate for hard partons dedends neither
on the momentum of the incomming parton nor on the number of flavors in the
QGP.

While $\Gamma _q^l$ is finite, $\Gamma _q^t$ is logarithmically infrared
divergent, as opposed to the quadratic singularity in (\ref{eq3})
resulting from Fig.1. The remaining divergence in the transverse part comes
from the absence of static magnetic screening in the effective gluon
propagator. A magnetic screening mass may arise at the nonperturbative
scale $g^2T$ as a result of possible magnetic monopole configurations
in the QGP \cite{LIN}. It is interesting to note that lattice \cite{DGT}
and classical monopole gas
\cite{BMA} calculations led to the same estimate,
$m_{mag}^2 \simeq 15\, \alpha _s^2\, T^2$. Another way of regularizing
this infrared singularity in a physical way relies in the self consistent
use of the quark damping in the quark propagator \cite{LSA,APG}, which
might be justified because, due to continuous scattering events in the
QGP, the quark will never be on mass shell exactly. The latter
regularization process also works in a QED gas, where magnetic monopole
configurations are absent. Including the both mechanisms phenomenologically
into $\Gamma _q^t$ \cite{PIB}, we end up with \cite{THB}
\begin{eqnarray}
\Gamma _q & \simeq & 1.3 \; \alpha _s\, T, \nonumber \\
\Gamma _g & \simeq & 3.0 \, \alpha _s\, T, \nonumber \\
\label{eq5}
\end{eqnarray}
where $\Gamma _g= (9/4)\, \Gamma _q$ is the interaction rate of a hard gluon.

Assuming these rates to be responsible for the thermalization of a parton
gas in ultrarelativistic heavy ion collisions yields thermalization times of
$\tau _g \simeq 0.7$ fm/c for gluons and $\tau _q \simeq 1.7$ fm/c for
quarks for typical values of $\alpha _s = 0.3$ and $T = 300$ MeV. These
results suggest a rapid thermalization of the gluon component and a
somewhat delayed of the complete parton gas (two-stage equilibration
\cite{SHA}).

Similar interaction rates are obtained using naive pQCD with the Debye
mass, $\mu _D^2=3\, m_g^2$, as an infrared regulator in the cross section.
For $N_F =2$ -- in this approximation the interaction rates depend
weakly on the number of flavors -- $\Gamma _q \simeq 1.1\, \alpha _s\,
T$ and $\Gamma _g \simeq 2.5\, \alpha _s\, T$ have been found \cite{THB}
in good agreement with the complete result (\ref{eq5}).

An important application of the ordinary gluon interaction rate is the
static color conductivity \cite{SGA}, which becomes $\sigma _c \simeq
2.4\, T$ using the result for $\Gamma _g$ quoted above in (\ref{eq5}).
Furthermore,
the ordinary interaction rate serves as an input for the collisional
energy loss of a high energy parton in the QGP \cite{BTB,THA}.

\yoursection{TRANSPORT RATES}

In a plasma with long range interactions, as for instance the QGP, the
transport interaction rate
\begin{equation}
\Gamma _{trans} = \int d\Gamma \> {\sin ^2\theta \over 2}
\label{eq6}
\end{equation}
describing momentum relaxation, (in contrast to color relaxation), is
in general of greater physical significance than the ordinary rate.
The transport weight $\sin^2\theta $, containing the scattering angle
$\theta $ in the center of mass system, changes the infrared behavior
of the rate significantly. Because of $\sin ^2\theta \sim q^2$ for small
momentum transfers $q$, the transport rate turns out to be only
logarithmically infrared divergent in naive perturbation theory, in contrast
to (\ref{eq3}),
\begin{equation}
\Gamma _{trans} \sim \alpha _s\> \int {dq\over q}
\label{eq7}
\end{equation}
but finite using the Braaten-Pisarski method due to dynamical screening
\cite{PBM}. Such quantities can be calculated consistently by introducing
a separation scale $q^\star $, as discussed by Braaten and Yuan \cite{BRY}.
The soft contribution,
according to momentum transfers $q<q^\star $, is obtained by using the
effective gluon propagator, while the hard contribution ($q>q^\star $)
follows from utilizing the bare propagator. Employing the restriction
$gT\ll q^\star \ll T$ and adding up the soft and hard parts, the arbitrary
separation scale drops out. Another example for a quantity calculable
in this way is the energy loss \cite{BTA}.

In the case of the ordinary
interaction rate the hard part only contributes to higher order in $g$
because this rate is dominated by soft momentum transfers, as opposed
to the transport rate, as can be seen by comparing (\ref{eq3}) and
(\ref{eq7}). The hard part requires to consider not only the
$t$-channel scattering diagram (Fig.1a) but all the others $2\rightarrow 2$
processes to lowest order \cite{CKR,CUS}.

In contrast to the the ordinary rate, the transport rate depends on the
momentum as well as $N_f$.
Using the Braaten-Yuan prescription, the following results for thermal
partons ($\langle p\rangle \simeq 3T$) have been obtained \cite{THB}
\begin{eqnarray}
\Gamma _{g,trans}(N_f=0) & \simeq & 5.2\, \alpha _s^2\, T\>
\ln {0.25\over \alpha _s},\nonumber \\
\Gamma _{g,trans}(N_f=2) & \simeq & 6.6\, \alpha _s^2\, T\>
\ln {0.19\over \alpha _s},\nonumber \\
\Gamma _{q,trans}(N_f=2) & \simeq & 2.5\, \alpha _s^2\, T\>
\ln {0.21\over \alpha _s}. \nonumber \\
\label{eq8}
\end{eqnarray}
It should be noted that, owing to the different infrared behavior, the
transport rate is reduced by one order of $\alpha _s $ compared to the
ordinary rate (\ref{eq5}) corresponding to a much larger thermalization
time in the weak coupling limit than following from (\ref{eq5}). Furthermore,
the factor $\ln (const/\alpha _s)$
in the transport rate originates from a sensitivity to the scale $gT$,
which allows a calculation of the constant under the logarithm, i.e.
beyond the leading logarithm approximation, in contrast to the
logarithmic factor in the ordinary rate coming from a sensitivity to $g^2T$,
which cannot be treated by the Braaten-Pisarski method.

Unfortunately, the transport rate becomes negative for $\alpha _s
{\buildrel >\over \sim }0.2 $, indicating the breakdown of pQCD for
realistic values of the coupling constant, as discussed in the next
section, and rendering an estimate of the thermalization times impossible.

The result (\ref{eq8}) may be compared to the simple approximation
of using a bare gluon propagator and cutting off the integral
by the Debye mass. In the pure gluonic case, we then have
only to substitute the constant under the logarithm 0.25 in (\ref{eq8})
by 0.33 \cite{THB} showing again the approximate validity of this
simplified method.

The shear viscosity coefficient of the QGP beyond the relaxation time
approximation is related to the transport rate and the energy
density of the QGP $\epsilon $ by $\eta = 2\epsilon /(15 \Gamma _{trans})$
\cite{DAG,REI,THC}, which leads with (\ref{eq8}) to
\begin{equation}
\eta = {T^3\over \alpha _s^2}\> \left [{0.11\over \ln
(0.19/\alpha _s )}+{0.37\over \ln (0.21/\alpha _s)}\right ].
\label{eq9}
\end{equation}
In Fig.3 the ratio $\eta /T^3$ is shown as a function of $\alpha _s$ and
compared to a result obtained within the leading logarithm approximation
\cite{BMP}. The horizontal line gives the upper limit for the validity
of the Navier-Stokes equation \cite{DAG}, indicating the importance of
dissipative effects in the QGP.

\begin{figure}[htp]
\vskip-0.5cm
\centerline{\psfig{figure=lbl3.ps,height=10.0cm}}
\vskip-0.5cm
\caption{Shear viscosity coefficient according to (9) (solid line)
and Ref.39 (dashed line). The dotted line indicates the upper
limit for the validity of the Navier-Stokes equation [11].}
\end{figure}

\yoursection{LIMITS OF PERTURBATIVE QCD}

In this section, we discuss the range of validity of pQCD for
investigating properties of the QGP in relativistic heavy ion collisions.
For this purpose, we consider various quantities computed within pQCD
so far. We distinguish three classes of quantities.

The first class consists of quantities which are finite and gauge
invariant using naive perturbation theory. The gluon and quark self
energies in the high temperature limit, which is equivalent to the
hard thermal loop approximation, belong to this class [40-42].
 From these we get the dispersion relations of the collective modes
(plasmons and plasminos) in the QGP as well as the plasma frequency
$m_g$, the Debye mass $\mu _D$, and the effective quark mass $m_q =
gT/{\sqrt 6}$ to leading order in $g$. These results should hold, as long
as $\alpha _s$ is small, which is the condition for the validity of
naive pQCD.

The second class contains quantities which are logarithmically infrared
divergent in naive perturbation theory but finite using the effective one,
as for example the transport rate and the energy loss. The transport rate
for thermal partons becomes unphysical, i.e. negative, if the strong
coupling constant exceeds $\alpha _s \simeq 0.2$ (see (\ref{eq8}))
corresponding to $g \simeq 1.6$. For high energy partons, $p\gg T$,
on the other hand, the rate is positive even for values of $\alpha _s
\simeq 0.5$ because of the factor $\ln (s/{q^\star} ^2)$ in the hard
contribution \cite{THB}, where $\sqrt s $ is the total energy in the
center of mass system. The same behavior has been observed in the case
of the energy loss of a heavy quark \cite{BTB}, where for low momenta
above the thermal one and $g \geq 1.1$ an unphysical negative result occurs,
whereas for high momenta the calculation appears to be reliable.

 From Fig.3 we see that the perturbative approximation seems to hold even
close to the critical value of the coupling constant before the curves
bend upwards. Thus we speculate that quantities of the second class,
containing important observables, can be calculated consistently within
leading order pQCD only as long as $g {\buildrel <\over \sim } 1$ or
$p\gg T$, as sketched in Fig.4. The failure of extrapolating perturbative
results at thermal momenta above $g \simeq 1$ is
not surprising since at this value the distinction between the scales $T$,
$gT$, and $g^2T$ becomes meaningless. On the contrary, considering $g\ll 1$
as a necessary prerequisite for the effective perturbation theory, the
Braaten-Pisarski method works for astonishingly large values of $g$.
It should
be noted here, however, that a recent computation of the plasma frequency
at next to leading order, which also belongs to the second class, breaks
down not till $g \geq 3.2$ \cite{SCA}.

\begin{figure}[htp]
\vskip-0.5cm
\centerline{\psfig{figure=lbl4.ps,height=10.0cm}}
\caption{The grey area roughly excludes the range of validity of pQCD.}
\end{figure}

The third class comprises quantities which are quadratically infrared
divergent in naive and logarithmically in effective perturbation theory,
as e.g. the ordinary interaction rate (\ref{eq4}) and the Debye mass
beyond the leading order \cite{REB}. These quantites cannot be calculated
unambiguously so far within pQCD requiring the developement of techniques
beyond the Braaten-Pisarki method.

Finally, we will discuss why a negative result for the transport rate,
which follows from the square of the magnitude of a matrix element, can
arise and how this situation can be improved in principle. For this purpose,
we will exemplify the problem replacing the complicated effective gluon
propagator by a simplified version
\begin{equation}
\Delta (t) = {1\over t+\mu ^2},
\label{eq10}
\end{equation}
where $\mu \sim g$ and $t=-(P-P')^2$. Also, instead of the transport rate, we
consider the integral
\begin{equation}
\gamma = g^4\> \int _0^s dt\> t\> |\Delta (t)|^2.
\label{eq11}
\end{equation}
The factor $t$ in front of $|\Delta (t)|^2$ comes from the transport weight
$\sin ^2\theta $ \cite{THB}.

Introducing a separation scale $t^\star $ according to the Braaten-Yuan
method and using the bare
propagator $\Delta _0(t) = 1/t$ for $t>t^\star $ and $\Delta (t)$ for
$t<t^\star $, we get
\begin{eqnarray}
\gamma (t>t^\star ) & = & g^4\> \ln {s\over t^\star },\nonumber \\
\gamma (t<t^\star ) & = & g^4\> \left (\ln {t^\star +\mu ^2\over \mu ^2}
+{\mu ^2\over t^\star +\mu ^2}-1\right ).\nonumber \\
\label{eq12}
\end{eqnarray}
Assuming $t^\star \gg \mu ^2$ in the soft part, the separation
scale drops out after adding the both contributions, leading to
\begin{equation}
\gamma = g^4\> \left ( \ln {s\over \mu ^2}-1 \right ).
\label{eq13}
\end{equation}
Now, if $g$ is of the order of one and $s$ of the order of $T^2$, we
get into trouble, because $\gamma <0$ is possible. The reason for
this unphysical behavior is that we neglected positive terms in the
soft part or had to assume $t^\star >s$ in the hard part.

The problem can be circumvented by using the full propagator over
the entire integration range
\begin{equation}
\gamma  =  g^4\> \left (\ln {s +\mu ^2\over \mu ^2}
+{\mu ^2\over s+\mu ^2}-1\right ),
\label{eq14}
\end{equation}
which is always positive. However, now $\gamma $ contains higher
orders in $g$, since $\mu \sim g$, and cannot be written in the form
\begin{equation}
\gamma \sim g^4\> \ln {const \over g^2},
\label{eq15}
\end{equation}
where the constant is independent of $g$. It is the purpose of the
Braaten-Yuan Method to isolate the leading order consistently, whereas
the result using the effective gluon propagator over the entire
momentum range is contaminated by higher orders. Furthermore, the latter
result is not complete to the order of the coupling constant involved,
because additional diagrams, containing e.g. effective vertices, will
contribute to the same order beyond the leading one. Even worse, the
next to leading order contributions are expected to show a sensitivity to
the scale $g^2T$ resulting in infrared singularities. Thus the requirement
for gauge invariance and completeness in the coupling constant, fulfilled
by using the Braaten-Yuan method, which is well defined in the weak
coupling limit, leads to negative unphysical results
to leading order if the coupling constant exceeds a critical value.
In order to cure this problem, one has to go beyond the leading order
approximation increasing the complexity of the calculation enormously
and encountering infrared divergences presumably.

\yoursection{CONCLUSIONS}

The parton interaction rates are important quantities which are closely
related to interesting observables of the QGP, as thermalization times,
mean free paths, energy losses, shear viscosity, and color conductivity.
The ordinary interaction rate addressed by pQCD suffers from a
logarithmic infrared singularity even using the resummation technique
by Braaten and Pisarski. Applying realistic cutoffs, interaction rates
for quarks $\Gamma _q \simeq 1.3\, \alpha _s\, T$ and for gluons
$\Gamma _g \simeq 3\, \alpha _s\, T$, respectively, have been estimated.

Considering thermalization times, one should rather study transport
rates which describe momentum relaxation via a transport weight
$\sin ^2\theta $. This weight changes the infrared behavior of the rates
completely, leading to a finite result within the Braaten-Pisarski method,
which is one order of $\alpha _s$ higher compared to the ordinary rate.
Beyond the leading logarithm approximation, we found in the case of two
active flavors in the QGP  for thermal momenta
$\Gamma ^q_{trans} \simeq 2.5\, \alpha _s^2\, \ln (0.21/\alpha _s)$
and $\Gamma ^g_{trans} \simeq 6.6\, \alpha _s^2\, \ln (0.19/\alpha _s)$.
Also the shear viscosity coefficient (\ref{eq9})
follows from the transport rate.

Unfortunately, an extrapolation to realistic values of the coupling
constant $\alpha _s \simeq 0.2$ -- 0.5 is not possible since the
transport rates for thermal partons become negative indicating a breakdown
of pQCD to leading order. A similar behavior has been observed in the case
of the energy loss. This leads us to the speculation that the important
class of quantities which are logarithmically infrared divergent in
naive perturbation theory can be calculated consistently to leading
order only for $\alpha _s {\buildrel <\over \sim }0.1$.  In order to
obtain meaningful results above this critical coupling constant,
one has to go at least to the next to leading order.

Finally, in order to obtain a satisfactory estimate for many purposes,
the simplification of using naive pQCD including the Debye mass
as an infrared regulator is sufficient in the case of the ordinary
as well as the transport interaction rate.

\yoursection{Acknowledgments}

Valuable discussions with T.S. Bir\'o and H. Heiselberg are acknowledged.
The work was supported by the BMFT and GSI Darmstadt.

\end{document}